\documentclass{article}
\usepackage{spconf,amsmath,graphicx,amsfonts,hyperref}
\usepackage{multirow}

\title{Unsupervised Learning of Semantic Audio Representations}
\makeatletter
\def\@name{ \emph{Aren Jansen, Manoj Plakal, Ratheet Pandya, Daniel
    P. W. Ellis,\thanks{Portions of this paper were submitted to ML4Audio 2017 workshop.} } \\
  \emph{Shawn Hershey, Jiayang Liu, R. Channing Moore, Rif A. Saurous}\vspace{6pt}}
\makeatother
\address{Google, Inc., Mountain View, CA, and New York, NY, USA\\
{\footnotesize \tt\{arenjansen,plakal,ratheet,dpwe,shershey,jiayl,channingmoore,rif\}@google.com}}

\begin{document}
\ninept
\maketitle
\begin{abstract}
Even in the absence of any explicit semantic annotation, vast collections of
audio recordings provide valuable information for learning the categorical
structure of sounds.  We consider several class-agnostic semantic constraints
that apply to unlabeled nonspeech audio: (i) noise and translations in time do
not change the underlying sound category, (ii) a mixture of two sound events
inherits the categories of the constituents, and (iii) the categories of events
in close temporal proximity are likely to be the same or related.  Without
labels to ground them, these constraints are incompatible with classification
loss functions. However, they may still be leveraged to identify geometric
inequalities needed for triplet loss-based training of convolutional neural
networks.  The result is low-dimensional embeddings of the input spectrograms
that recover 41\% and 84\% of the performance of their fully-supervised
counterparts when applied to downstream query-by-example sound retrieval and
sound event classification tasks, respectively.  Moreover, in
limited-supervision settings, our unsupervised embeddings double the
state-of-the-art classification performance.
\end{abstract}


\section{Introduction}
\label{sec:intro}

The last few years have seen great advances in nonspeech audio processing, as
popular deep learning architectures developed in the speech and image processing
communities have been ported to this relatively understudied
domain~\cite{takahashi2016deep,hershey2017cnn,cakir2017convolutional,wang2017first}.
However, these data-hungry neural networks are not always matched to the
available training data in the audio domain. While unlabeled audio is easy to
collect, manually labeling data for each new sound application remains
notoriously costly and time consuming.  We seek to alleviate this incongruity by
developing alternative learning strategies that exploit basic semantic
properties of sound that are not grounded to an explicit labeling.

Recent efforts in the computer vision community have identified several
class-independent constraints on natural images and videos that can be used to
learn semantic representations~\cite{zhang2016colorful}.  For example, object
categories are invariant to camera angle, and tracking unknown objects in videos
can provide novel examples for the same unknown
category~\cite{wang2015unsupervised}.  For audio, we can identify several
analogous constraints, which are not tied to any particular inventory of sound
categories.  First, we can apply category-preserving transformations to
individual events of unknown type, such as adding Gaussian noise, translation in
time within the analysis window, and small perturbations in frequency.  Second,
pairs of unknown sounds can be mixed to provide new, often natural sounding
examples of both.  Finally, sounds from within the same vicinity in a recording
likely contain multiple examples of the same (or related) unknown categories.

If provided a set of labeled sound events of the form \emph{$X$ is an example of
  category $C$}, applying the above semantic constraints is interpretable as
regular labeled data augmentation.  However, when each example has unknown
categorical assignment, standard classification loss functions can not be
applied.  We instead resort to deep metric learning using triplet
loss~\cite{weinberger2009distance,wang2014learning,hoffer2015deep}, which finds
a nonlinear mapping into a low dimensional space where simple Euclidean distance
can express any desired relationship between examples of the form \emph{$X$ is
  more like $Y$ than like $Z$}.  Critically, while labeled examples can be
converted to triplets to explicitly learn a semantic embedding (i.e., \emph{$X$
  has same class as $Y$, but different than $Z$}), a triplet relationship need
not be anchored to an explicit categorical assignment.  This makes it a natural
fit for our set of semantic constraints; indeed, a noisy version of a sound
event is more semantically similar to the clean recording than another arbitrary
sound.  Moreover, since we can generate as many triplets from as much unlabeled
data as we wish, we can support arbitrarily complex neural architectures.

To validate these ideas, we train embeddings using state-of-the-art
convolutional architectures on millions of triplets sampled from the
\emph{AudioSet} dataset~\cite{audioset}, both with and without using the label
information.  We evaluate the learned embeddings as features for
query-by-example sound retrieval and supervised sound event classification.  Our
results demonstrate that highly complex models can be trained from unlabeled
triplets alone to produce representations that recover up to 84\% of the
performance gap between using the raw log mel spectrogram inputs and using
fully-supervised embeddings trained on millions of labeled examples.

\section{Related Work}

There have been multiple past efforts to perform unsupervised deep
representation learning on non-speech audio.  Lee et
al.~\cite{lee2009unsupervised} applied convolutional deep belief networks to
extract a representation for speech and music, but not general purpose nonspeech
audio.  More recently, a denoising autoencoder variant was used to extract
features for environmental sound classification~\cite{xu2017unsupervised}.
While both approaches produced useful representations for their respective
tasks, neither explicitly introduced training mechanisms to elicit semantic
structure in their learned embeddings. Classical distance metric learning has
also been applied to music in the past~\cite{slaney2008learning}.

Recent zero-resource efforts in the speech processing community have explicitly
aimed to learn meaningful linguistic units from untranscribed
speech~\cite{versteegh2015zero}.  With this goal, several weak supervision
mechanisms have been proposed that are analogous to what we attempt to achieve
for nonspeech audio.  For speech, the relevant constraints are derived from the
inherent linguistic hierarchy: repeated unknown words have the same unknown
phonetic structure~\cite{jansen2013weak}, conversations with same unknown topic
have shared unknown words~\cite{frank2014weak}, etc.  Various forms of deep
metric
learning~\cite{synnaeve2014phonetics,kamper2015unsupervised,zeghidour2016joint,kamper2016deep}
have been successfully applied using these speech-specific constraints.

Finally, so-called self-supervised approaches in the computer vision community
are analogous to what we propose in this paper for audio.  There, constraints
based on egomotion~\cite{agrawal2015learning}, spatial/compositional
context~\cite{pathak2016context,doersch2015unsupervised}, object
tracking~\cite{wang2015unsupervised}, and colorization~\cite{zhang2016colorful}
have all been evaluated.  Recent efforts have extended this principle of
self-supervision to joint audio-visual models that learn speech or audio
embeddings using semantic constraints imposed by the companion visual
signal~\cite{arandjelovic2017look,aytar2016soundnet,harwath2016unsupervised,kamper2017visually}.

\section{Learning Algorithm}
\label{sec:alg}

Our training procedure consists of two stages: (i) sampling training triplets
from a collection of unlabeled audio recordings, and (ii) learning a map from
input context windows extracted from spectrograms (matrices with $F$
frequency channels and $T$ frames) to a lower $d$-dimensional vector space using
triplet loss optimization of convolutional neural networks. We summarize the
triplet loss metric learning framework, and then formally define each of our
triplet sampling strategies.

\subsection{Metric Learning with Triplet Loss}

The goal of triplet loss-based metric learning is to estimate a map $g:
\mathbb{R}^{F\times T} \rightarrow \mathbb{R}^d$ such that simple (e.g.)
Euclidean distance in the target space correspond to highly complex geometric
relationships in the input space.  Training data is provided as a set
$\mathcal{T} = \{t_i\}_{i=1}^N$ of example triplets of the form
$t_i\!=\!(x_a^{(i)},x_p^{(i)},x_n^{(i)})$, where $x_a^{(i)}, x_p^{(i)},
  x_n^{(i)}\!\in\!\mathbb{R}^{F\times T}$ are commonly referred to as the anchor,
  positive, and negative, respectively. The loss is given by

\begin{equation}
  \mathcal{L}(\mathcal{T}) = \sum_{i=1}^N \left[ \|g(x_a^{(i)})\!-\!g(x_p^{(i)})\|_2^2 -
    \|g(x_a^{(i)})\!-\!g(x_n^{(i)})\|_2^2 + \delta \right]_+,
  \label{eq:loss}
\end{equation}

\noindent
where $\|\!\cdot\!\|_2$ is $L_2$ norm, $[\cdot]_+$ is standard hinge loss, and
$\delta$ is a nonnegative margin hyperparameter.  Intuitively, the optimization
is attempting to learn an embedding of the input data such that positive
examples end up closer to their anchors than the corresponding negatives do, by
some margin.  Notice that the loss is identically zero when all training
triplets satisfy the inequality (dropping index)

\begin{equation}
  \|g(x_a)-g(x_p)\|_2^2 + \delta \leq \|g(x_a)-g(x_n)\|_2^2.
  \label{eq:ineq}
\end{equation}

\noindent
Thus we may also view the triplets as a collection of hard constraints on the
inputs.  This is an extremely flexible construct: any pairwise relationship
between input examples that permits a relative ranking (i.e., $(x_a,x_p)$ are
more similar than $(x_a,x_n)$) complies.  The learned distance then becomes a
proxy for that pairwise relationship.

The map $g$ can be defined by a fully-connected, $d$-unit output layer of any
modern deep learning architecture.  The optimization is performed with
stochastic gradient descent, though training time is greatly decreased with the
use of within-batch semi-hard negative mining~\cite{schroff2015facenet}.  Here,
all examples in the batch are transformed under the current state of $g$, and
the available negatives are reassigned to the anchor-positive pairs to make more
difficult triplets.  Specifically, we choose the closest negative to the anchor
that is still further away than the positive (the absolute closest is vulnerable
to label noise).


\subsection{Triplet Sampling Methods}
\label{sec:sampling}

\subsubsection{Explicitly Labeled Data}
In standard supervised learning, we are provided a set of labeled examples of
the form $\mathcal{Z}=\{(x_i,y_i)\}$, where each $x_i \in \mathbb{R}^{F\times
  T}$ and $y_i \in \mathcal{C}$ for some set $\mathcal{C}$ of semantic
categories.  Triplet loss-based metric learning was originally formulated for
this setting, and converting $\mathcal{Z}$ to a set of triplets is
straightforward.  For each $c \in \mathcal{C}$, we randomly sample
anchor-positive pairs $(x_a,x_p)$ from $\mathcal{Z}$ such that $y_a=y_p=c$.
Then, for each sampled pair, we attach as the triplet's negative an example
$(x_n,y_n) \in \mathcal{Z}$ such that $y_n \neq c$. This procedure sets the
supervised performance topline in our experiments.

\subsubsection{Gaussian Noise}
Since the introduction of the denoising autoencoders, learning representations
that are invariant to small perturbations in the original data space has been a
standard tool for unsupervised learning.  However, when more complex
convolutional architectures with pooling are desired, inverting the encoder
function is complicated~\cite{masci2011stacked}.  However, since we are not
interested in actually reconstructing the inputs, we can use triplet loss to
effect similar representational properties using an arbitrary deep learning
architecture.  For each $x_i\in\mathbb{R}^{F\times T}$ in the provided set of
unlabeled examples $\mathcal{X}$, we simply sample one or more anchor-positive
pairs of the form $(x_i,x_p)$, where element $x_{p,tf} = x_{i,tf}(1+|\epsilon_{tf}|)$
for $\epsilon_{tf} \sim \mathcal{N}(0,\sigma^2)$, a Gaussian distribution with mean 0
and standard deviation $\sigma$ (a model hyperparameter).  For each sampled
anchor-positive pair, we simply choose another example in $X$ as the negative to
complete the triplet.

\subsubsection{Time and Frequency Translation}
When processing long context windows of spectrogram, we are provided snapshots
of the contained sound events with arbitrary temporal offsets and clipping.  A
transient sound event with unknown category that starts at the left edge of the
window maintains its semantic assignment if it begins somewhere in the center.
However, in the input space, this simple translation in time can produce
dramatic transformations of the training data.  Similarly, small translations in
frequency may leave the semantics unchanged while greatly perturbing the input
space.  To exploit this we generate training triplets as follows. For each
$x_i\in\mathbb{R}^{F\times T}$ in the provided set of unlabeled examples
$\mathcal{X}$, we sample one or more anchor-positive pairs of the form
$(x_i,x_p)$ where $x_p = \mathrm{Trunc}_S(\mathrm{Circ}_T(x_i))$.  Here,
$\mathrm{Circ}_T$ is a circular shift in time by an integer number of frames
sampled uniformly from $[0,T-1]$, where $T$ is the number of frames in the
example. $\mathrm{Trunc}_S$ is a truncated shift in frequency by an integer
number of bins sampled uniformly from $[-S, S]$ (missing values after shift are
set to zero energy).  We again choose another example in $X$ as the negative to
complete the triplet.

\subsubsection{Example Mixing}
Sound is often referred to as transparent, since we can superimpose sound
recordings and still hear the constituents to some degree.  We can use this
intuition to construct triplets by mixing sounds together.  One approach to this
is forming a positive by mixing a random example with the anchor.  However, if
we subsequently attach a random negative, we can not guarantee we want to
satisfy the inequality of Eq.~\ref{eq:ineq}. For example, if the anchor is a dog
bark, and we mix it with a siren, we cannot assume a dog growl as the negative
should be mapped further from the anchor than the mixture.  To solve this, we
can mix random anchor and negative examples to form the positives. Given a
random anchor $x_a$ and negative $x_n$ containing energies in each
time-frequency cell, we construct positive $x_p = x_a + \alpha [E(x_a)/E(x_n)]
x_n$, where $E(x)$ is the total energy of example $x$, and $\alpha$ is a
hyperparameter.  Note that this is the only triplet type considered not strictly
compatible with semi-hard negative mining, since negatives are not randomly
sampled.

\subsubsection{Temporal Proximity}
Audio recordings from real world environments do not consist of events drawn
completely at random from the space of all possible sounds.  Instead, a given
environment has limited subset of sound creating objects that are often closely,
or even causally, related.  As such, two events in the same recording more
likely to be of the same, or at least related, event categories than any two
random events in a large audio collection.  We can use this intuition to sample
triplets of the form $(x_a,x_p,x_n)$ where $x_a$ and $x_p$ are from the same
recording, $x_n$ is from a different recording. We can further impose the
constraint that $|\mathrm{time}(x_a)-\mathrm{time}(x_p)| < \Delta t$, where
$\mathrm{time}(x)$ is the start time of example $x$, and $\Delta t$ is a
hyperparameter.  Note that if overlapping context windows and sufficiently small
values of $\Delta t$ are used, this method is functionally similar to the time
translation approach.

\subsubsection{Joint Training}
In supervised learning settings, it is not always trivial to combine multiple
data sources from separate domains with distinct label inventories.  For cases
where simply using a classification layer that is the union of the categories is
not possible (e.g. mutual exclusivity is not implied across the two class sets),
we must resort to multi-task training objectives.  For triplet loss, this is not
a problem.  All triplet sets produced using the sampling methods outlined in
this section can be simply mixed together for training a joint embedding that
reflects them all to whatever degree possible.  In general, if we have
preconceived notions of each constraint's importance, we can either introduce a
source-dependent weight to each triplet's contribution to the loss function in
Eq.~\ref{eq:loss} or, alternatively, use varying triplet sample sizes for each
source.

\section{Experiments}
\label{sec:exp}

We evaluate embeddings that result from the triplet sampling methods of
Section~\ref{sec:sampling} in two downstream tasks: (i) query-by-example
semantic retrieval of sound segments, and (ii) training shallow fully-connected
sound event classifiers.  The query-by-example task does not involve any
subsequent supervised training and thus directly measures the intrinsic semantic
consistency of the learned representation.  The shallow model measures how
easily a relatively simple, non-convolutional classifier network can predict the
sound event categories given the labeled data.  Finally, we also perform a
lightly-supervised classification experiment, where we repeat the shallow model
evaluation with only a small fraction of the labeled data.  This allows us to
measure the utility of unlabeled data in reducing annotation requirements for
any sound event classification application where unlabeled data is plentiful.   

\subsection{Dataset and Features}

We use Google's recently released \emph{AudioSet} database of manually annotated
sound events~\cite{audioset} for both training and evaluation. \emph{AudioSet}
consists of over 2 million 10-second audio segments from YouTube videos, each
labeled using a comprehensive ontology of 527 sound event categories (minimum
120 segments/class). We use an internal version of the unbalanced training set
(50\%-larger than the released set), which we split into train and development
subsets.  We report all performance metrics on the released evaluation set.  We
compute 64-channel mel-scale spectrograms using an FFT window size of 25 ms with
10 ms step.  Triplets are sampled in this energy domain since the some of our
triplet sampling mechanisms require an energy interpretation, but a stabilized
logarithm is applied before input to our models.  We then process these
spectrograms into non-overlapping 0.96 second context windows, such that each
training example is an $F$=64 by $T$=96 matrix. Each embedding model was trained
on order 10 million triplets (40 million for joint model).

\subsection{Model Architecture}

Given its impressive performance on previous large-scale sound classification
evaluations~\cite{hershey2017cnn}, we use the ResNet-50 convolutional neural
network architecture.  Each input $64\!\times\!96$ context window is first
processed by a layer of 64 convolutional $7\!\times\!7$ filters, followed by a
$3\!\times\!3$ max pool with stride 2 in both dimensions.  This is followed by 4
standard ResNet blocks and a final average pool over time and frequency to a
2048-dimensional representation.  Instead of the classification output layer
used in~\cite{hershey2017cnn}, all of our triplet models use a 128-unit
fully-connected linear output layer. This produces a vector of dimension
$d=$128, which represents a factor of 48 reduction from the original input
dimensionality of $64\!\times\!96$. We also employ the standard practice of
length normalizing the network output before input to the loss function (i.e.,
$g = h/\|h\|_2$, where $h$ is the output embedding layer of the network). This
normalization means the squared Euclidean distance used in the loss is
proportional to using cosine distance, a common choice for learned
representations. All training is performed using Adam, with hyperparameters
optimized on the development set.  We use semi-hard negative mining and a
learning rate of $10^{-4}$ for supervised, temporal proximity, and combined
unsupervised models; otherwise $10^{-6}$ is used with mining disabled.  The
margin hyperparameter $\delta$ is set to 0.1 in all cases.

\subsection{Query-by-Example Retrieval}

\begin{table}
\vspace{-0.2cm}
  \centering
  \caption{Segment retrieval mean average precision (mAP) as function of: (left)
    Gaussian width $\sigma$ for Gaussian noise triplets; (middle) frequency
    shift range $S$ for translation triplets; (right) and mixing weight $\alpha$
    for mixed example triplets. Best results in bold.}
  \label{tab:hyperparam}
  \vspace{0.2cm}
  \begin{tabular}{c|c}
    \hline \hline
    $\sigma$ & mAP \\
    \hline \hline
    0.1 & 0.453 \\
    0.25 & 0.466 \\
    0.5 & {\bf 0.478}\\
    1.0 & {\bf 0.478} \\
    \hline \hline
  \end{tabular}
  \hspace{0.5cm}
  \begin{tabular}{c|c}
    \hline \hline
    $S$ & mAP \\
    \hline \hline
    0 & 0.461\\
    2 & 0.492 \\
    5 & 0.493 \\
    10 & {\bf 0.508} \\
    \hline \hline
  \end{tabular}
  \hspace{0.5cm}
  \begin{tabular}{c|c}
    \hline \hline
    $\alpha$ & mAP \\
    \hline \hline
    0.1 & 0.483 \\
    0.25 & {\bf 0.489}\\
    0.5 & 0.487\\
    1.0 & 0.476\\
    \hline \hline
  \end{tabular}
\vspace{-0.2cm}
\end{table}

\begin{table*}[t]
\vspace{-0.2cm}
  \centering
  \caption{Mean average precision for segment retrieval and shallow model
    classification using original log mel spectrogram and triplet embeddings as
    features. All embedding models use the same ResNet-50 architecture with a
    128-dimensional linear output layer.}
  \label{tab:results}
  \vspace{0.2cm}
  \begin{tabular}{c|c|c|c|c|c|c}
    \hline \hline
    & \multicolumn{2}{c|}{} & \multicolumn{2}{c|}{Classification} & \multicolumn{2}{c}{Classification}\\
    & \multicolumn{2}{c|}{QbE Retrieval} & \multicolumn{2}{c|}{(1 layer, 512 units)} &
    \multicolumn{2}{c}{(2 layer, 512 units)}\\
    \cline{2-7}
    Representation & mAP & recovery & mAP & recovery & mAP & recovery\\
    \hline \hline
    Explicit Label Triplet (topline)  & 0.790 & \emph{100\%} & 0.288 & \emph{100\%} & 0.289 & \emph{100\%} \\
    Log Mel Spectrogram (baseline) & 0.423 & \emph{0\%}   & 0.065 & \emph{0\%}   & 0.102 &   \emph{0\%}\\
    \hline 
    Gaussian Noise ($\sigma\!=\!0.5$)  & 0.478 & 15\%         & 0.096 & 14\%  & 0.114 &   6\%\\
    T/F Translation ($S\!=\!10$)       & 0.508 & 23\%         & 0.108 & 19\%  & 0.125 &  12\%\\
    Mixed Example ($\alpha\!=\!0.25$)  & 0.489 & 18\%         & 0.103 & 17\%  & 0.122 &  11\%\\
    Temporal Proximity ($\Delta t\!=\!10 s$) & 0.562 & 38\%  & 0.226 & 72\%  & 0.241 &  74\%\\
    \hline 
    Joint Unsupervised Triplet         & 0.575 & 41\%         & 0.244 & 80\%  & 0.259 &  84\%\\
    \hline  \hline
  \end{tabular}
\vspace{-0.2cm}
\end{table*}

Our first evaluation task is query-by-example (QbE) segment retrieval.  Here, no
additional training is performed, making it a direct measurement of inherent
semantic representational quality.  We begin by mapping each 0.96 second context
window in the evaluation set to its corresponding 128-dimensional embedding
vector and average these across each \emph{AudioSet} segment to arrive at a
segment-level embedding.  For each sound event category, from the
\emph{AudioSet} evaluation set we sample 100 segments where it is present, and
100 segments where it is not.  We then compute the cosine distance between all
4,950 within-class pairs as target trials, and all 10,000 (present,not-present)
pairs as nontarget trials.  We sort this set of pairs by ascending distance and
compute the average precision (AP) of ranking target over nontarget trials
(random chance gives 0.33).  We repeat this for each class and average the
per-class AP score to produce the reported mean average precision (mAP).

Table~\ref{tab:hyperparam} shows the retrieval mAP for three of the triplet
sampling mechanisms as a function of their associated hyperparameters.   In each
case, the optimal performing setting on the validation set also was optimal for
eval (listed for each hyperparameter in bold).  For the Gaussian noise and
example mixing, we observed relatively weak dependence on sampling
hyperparameter values.  However, we found for the translation method that
allowing larger shifts in frequency produce substantial improvements over time
translations alone.  While this may be surprising from a signal processing point
of view, two-dimensional translations help to force increased spectral
localization in the early layers' filters of the convolutional network, which is
observed in fully-supervised models.  Note that since all \emph{AudioSet} clips
are limited to at most 10 seconds, we did not explore additional limiting of
proximity (i.e., $\Delta t$ is effectively clip duration).

Table~\ref{tab:results} shows the retrieval performance for each of the
evaluated representations.  The fully-supervised topline uses explicitly labeled
data to sample the triplets.  As a baseline, we evaluate the retrieval
performance achieved using the raw log mel spectrogram features (each
$64\!\times\!96$ context window is treated as a 6144-dimensional vector before
segment-level averaging).  For each of the unsupervised methods, we tuned on the
development set and the reported performance here is on the separate evaluation
set.  At the bottom, we also list the performance of the joint embedding,
trained on a mixture of all four unsupervised triplet types (approximately equal
number of triplets from each).  Alongside each mAP value, we also list the
percentage of the baseline-to-topline performance gap recovered using each given
unsupervised triplet embedding.  We find that each unsupervised triplet method
significant improves retrieval performance over the input features, with the
joint unsupervised model improving mAP by 15\% absolute over the input
spectrogram features, and recovering more than 40\% of the performance gap.

\begin{table}
\vspace{-0.2cm}
  \centering
  \caption{Lightly-supervised classifier performance averaged over three trials,
    each trained with a different random draw of 20 segments/class (totaling
    0.5\% of labeled data).}
  \label{tab:light}
    \vspace{0.2cm}
  \begin{tabular}{c|c|c}
    \hline  \hline
    Representation & Classifier Architecture & mAP \\
    \hline  \hline
    Log Mel Spectrogram & Fully Connected (4x512) & 0.032 \\
    Log Mel Spectrogram & ResNet-50               & 0.072 \\
    \hline
    Joint Unsupervised Triplet & Fully Connected (1x512) & 0.143 \\
    \hline  \hline
  \end{tabular}
\vspace{-0.2cm}
\end{table}

\subsection{Sound Classification}

While the retrieval task measures how the geometric structure of the
representation mirrors \emph{AudioSet} classes, we are also interested how our
unsupervised methods aid an arbitrary downstream supervised task over the same
or similar data.  To test this, we use our various embeddings to train shallow,
fully-connected networks using labeled \emph{AudioSet} segments.  For each
feature, we consider classifiers with 1 and 2 hidden layers of 512 units each.
The output layer consists of independent logistic regression models for the 527
classes.  For each class, we compute segment-level scores (average of the
frame-level predictions) for the evaluation set and compute average precision.
We again report the mean average precision over the classes.

Table~\ref{tab:results} shows the classification performance for each of
representation types. We again find substantial improvement over the input
features in all cases, with temporal proximity the clear standout.  Combining
triplet sets provides additional gains, indicating the learned representation's
ability to encode multiple types of semantic constraints for downstream tasks.
Notice that our approach performs \emph{fully-unsupervised} training of a
ResNet-50 triplet embedding model that achieves 85\% (0.244/0.288) the mAP of a
\emph{fully-supervised} ResNet-50 triplet embedding model, when both are coupled
to a single hidden layer downstream classifier.

Finally, Table~\ref{tab:light} shows performance of lightly-supervised
classifiers trained on just 20 examples per class.  To account for the
variability in sample selection, we generate 3 random training samples, run the
experiment separately on each, and report average performance.  Here we evaluate
three models: (i) a ResNet-50 classifier model, (ii) a fully-connected model
trained from log mel spectrograms, and (iii) a fully-connected model trained on
the joint unsupervised triplet embedding (last line of
Table~\ref{tab:results}). Since our unsupervised triplet embeddings are derived
from the full \emph{AudioSet} train set (as unlabeled data), a single layer
classifier trained on top doubles the mAP of a full ResNet-50 classifier trained
from raw inputs.


\section{Conclusions}

We have presented a new approach to unsupervised audio representation learning
that explicitly elicits semantic structure.  By sampling triplets using a
variety of audio-specific semantic constraints that do not require labeled data,
we learn a representation that greatly outperforms the raw inputs on both sound
event retrieval and classification task.  We found that the various semantic
constraints are complementary, producing improvements when combined to train a
joint triplet loss embedding model.  Finally, we demonstrated that our best
unsupervised embedding provides great advantage when training sound event
classifiers in limited supervision scenarios.

\clearpage
\bibliographystyle{IEEEbib}
\bibliography{icassp18}

\end{document}